\documentclass[useAMS,usegraphicx,usenatbib]{mn2e}
\usepackage{aas_macros}     

\title[The assembly bias of dark matter haloes to higher orders]
{The assembly bias of dark matter haloes to higher orders}
\author[Angulo  et al.]{
\parbox[h]{160mm}{
R. E. Angulo\thanks{E-mail: raul.angulo@durham.ac.uk},
C. M. Baugh\thanks{E-mail: c.m.baugh@durham.ac.uk},
C. G. Lacey\thanks{E-mail: cedric.lacey@durham.ac.uk},
}\vspace{6pt}\\
Institute for Computational Cosmology, Department of Physics, 
University of Durham, South Road, Durham, DH1 3LE, UK. 
\vspace*{-0.5cm}}

\begin{document}
\date{\today}
\pagerange{\pageref{firstpage}--\pageref{lastpage}} \pubyear{2007}
\maketitle
\label{firstpage}

\begin{abstract}
We use an extremely large volume ($2.4h^{-3}{\rm Gpc}^{3}$), high 
resolution N-body simulation to measure the higher order clustering 
of dark matter haloes as a function of mass and internal structure. 
As a result of the large simulation volume and the use of a novel 
``cross-moment'' counts-in-cells technique which suppresses 
discreteness noise, we are able to measure the clustering of haloes 
corresponding to rarer peaks than was possible in previous studies; 
the rarest haloes for which we measure the variance are 100 times 
more clustered than the dark matter. We are able to extract, 
for the first time, halo bias parameters from linear up to fourth order. 
For all orders measured, we find that the bias parameters are a 
strong function of mass for haloes more massive 
than the characteristic mass $M_{*}$. Currently, no theoretical model 
is able to reproduce this mass dependence closely. We find that 
the bias parameters also depend on the internal structure of the halo up to 
fourth order. For haloes more massive than $M_{*}$, we find that the more 
concentrated haloes are more weakly clustered than the less concentrated ones. 
We see no dependence of clustering on concentration for haloes with masses 
$M<M_{*}$; this is contrary to the trend reported in the literature 
when segregating haloes by their formation time. Our results 
are insensitive to whether haloes are labelled by the total mass 
returned by the friends-of-friends group finder or by the 
mass of the most massive substructure. This implies that our conclusions 
are not an artefact of the particular choice of group finding algorithm. 
Our results will provide important input to theoretical models of 
galaxy clustering. 
\end{abstract}

\section{Introduction}

The spatial distribution of dark matter haloes is 
not as simple as was once suspected. In the standard 
theoretical model for the abundance and distribution 
of haloes, the clustering strength of haloes is predicted to be a 
function of mass alone, with more massive haloes displaying 
stronger clustering (e.g. \citealt{Kaiser1984, ColeKaiser1989, MoWhite1996}). 
However, recent numerical simulations of 
hierarchical cosmologies, by covering larger volumes 
with ever improving mass resolution, have been able to reveal 
subtle dependences of halo clustering on other properties such as 
formation redshift, the internal structure of the halo and 
its spin (\citealt*{GaoSpringelWhite2005}; \citealt{Wechsler2006, Harker2006, Bett2007, Wetzel2007}; \citealt*{JingSutoMo2007};
\citealt{Espino-Briones2007}). 

The dependence of halo clustering on a second parameter in addition 
to mass is generally referred to as assembly bias. However, the 
nature of the trend in clustering strength  recovered depends upon 
the choice of property used to classify haloes of a given mass. 
Early simulation work failed to uncover a convincing assembly bias 
signal, as a result of insufficient volume and mass resolution, which 
meant that halo clustering could be measured for only a narrow range of mass 
and with limited statistics (\citealt{Lemson1999,Percival2003,ShethTormen2004}) 
The first clear indication of a dependence of halo clustering 
on a second property was uncovered by \cite*{GaoSpringelWhite2005}. 
These authors reported that low mass haloes which form early are more 
clustered than haloes of the same mass which form later on. No effect 
was seen for massive haloes. \cite{Wechsler2006} were able to confirm 
this result but also found that halo clustering depends on the density profile 
of the halo, as characterized by the concentration parameter \citep*{NFW1997}. 
The sense of the dependence of clustering strength on concentration changes with mass. 
Wechsler et~al. found that massive haloes showed a dependence of clustering strength 
on concentration, with low concentration haloes being the more strongly clustered 
(as confirmed by \citealt{GaoWhite2007}, \citealt*{JingSutoMo2007} and 
\citealt{Wetzel2007}). This trend of clustering 
strength with concentration is reversed 
for low mass haloes. Although formation time 
and concentration are correlated (e.g. \citealt{Neto2007}), their impact on 
the clustering of haloes does not follow trivially from this correlation, 
suggesting that some other parameter may be more fundamental (as argued by
\citealt*{CrotonGaoWhite2007}). 

Previous studies of assembly bias have focused exclusively on the linear 
bias parameter, which relates the two-point correlations of haloes and dark 
matter. Measurements from local surveys have shown that galaxies 
have significant higher order correlation functions and that the spatial 
distribution of galaxies and haloes is not fully described by two-point 
statistics (e.g. \citealt{2004MNRAS351L44B,2004MNRAS.352.1232C,2006MNRAS.368.1507N,2006MNRAS.373..759F}). 
With large surveys planned at higher redshifts, there is 
a clear need for accurate models of the higher order clustering of dark matter 
haloes, and to establish whether or not the higher order bias parameters depend 
on other properties in addition to mass.  

In this paper we measure the higher order bias parameters of dark matter haloes 
using a simulation which covers a volume more than an order of magnitude larger 
than the run analyzed by Gao and collaborators. We use a novel approach to estimate 
the higher order correlation functions of dark matter haloes. Our method builds upon  
the cross-correlation technique advocated for two-point correlations by \cite*{JingSutoMo2007},\cite{GaoWhite2007}
and \cite{Smith2007}. By considering fluctuations in the density of haloes and dark 
matter within the same smoothing window, we can suppress discreteness noise in our measurements. 
This improved clustering estimator, which uses the counts-in-cells method, when 
coupled with the large volume of our simulation, allows 
us to recover the bias parameters from linear to fourth order, and to study 
the dependence of these parameters on the halo concentration. 

In Section 2, we give the theoretical background to the counts-in-cells technique 
we use to estimate higher order clustering and explain how the clustering of haloes 
relates to the underlying dark matter at different orders. We also introduce the 
numerical simulations in that section. We present our results in Section 3 and 
a summary and discussion in Section 4.

\section{Theoretical background and method}

In this Section we give the theoretical background to the measurements presented in Section 3. 
We estimate the clustering of haloes and dark matter using a counts-in-cells approach. An overview 
of this method is given in \S 2.1, in which we explain how to obtain expressions for the higher 
order auto correlation functions of a density field from the moments of the distribution of 
counts-in-cells. We also introduce the concept of higher order {\it cross} correlation functions, 
which combine fluctuations in two density fields. The concept of hierarchical amplitudes, 
scaling relations between higher order correlation functions and the two-point correlation 
function, is introduced in \S 2.2. 
The key theoretical results relating the higher order cross correlation functions of haloes to  
the two-point function and hierarchical amplitudes of the dark matter are given in \S2.3. The 
simulations we use to measure the clustering of dark matter haloes are described in \S2.4.

\subsection{The counts in cells approach to measuring clustering}

Here we give a brief overview of the approach of using the distribution 
of counts in cells to estimate the higher order auto correlation functions 
of a set of objects. An excellent and comprehensive review of this material is given 
by \cite{Bernardeau2002}. We first discuss the higher order correlation 
functions for the case of a continuous, unsmoothed density field, then introduce 
the concept of cross-correlations (\S 2.1.1), before explaining how these 
results are changed in the case of a smoothed distribution of discrete points (\S 2.1.2).

\subsubsection{Higher order correlations: unsmoothed and continuous density field}

In general, the complete hierarchy of $N$-point correlation functions is required to 
fully characterize the spatial distribution of fluctuations in a density field. 
An exception to this occurs for the special case of a Gaussian density field, which  
can be described completely by its two-point correlation function. 

The $N$-point correlation functions are usually written in terms of the dimensionless 
density fluctuation or density contrast at a point: 
\begin{equation}
\delta(x) = \rho(x) / \langle \rho \rangle - 1,  
\end{equation}
where $\langle \rho \rangle$ is the mean density; the average is taken over different 
spatial locations. By definition, $\langle \delta(x) \rangle = 0$ when the average is 
taken over a fair sample of the density field. 
The $N^{\rm th}$ order moment of the density field, sometimes referred to as a central 
moment, because $\delta$ is a fractional fluctuation around the mean density, is 
given by: 
\begin{equation} 
\mu_{N} = \langle \delta(x_1), \ldots, \delta(x_n) \rangle, 
\label{eq:unconnected}
\end{equation}  
where, in general, the density fluctuations are correlated at different spatial 
locations. 

The $N^{\rm th}$ order central moments defined in Eq.~\ref{eq:unconnected} can be 
decomposed into terms which include products of lower order moments. This is because 
there are different permutations of how the $N$-points can be ``connected'' or joined 
together. This idea is illustrated nicely by tree diagrams in the review by 
\cite{Bernardeau2002}. The terms into which the central moments are broken down 
are called connected moments and these cannot be reduced further. In the tree diagram 
language, an $N$-point connected moment has no disjoint points; all $N$-points are 
linked to one another when the spatial averaging is performed. The distinction between 
connected and unconnected moments may become clearer if we write down the decomposition 
of the unconnected central moments up to fifth order: 

\begin{eqnarray} 
\langle \delta^{2} \rangle &=& \langle \delta^{2} \rangle_{c} + \langle \delta \rangle^{2}_{c} \\
\langle \delta^{3} \rangle &=& \langle \delta^{3} \rangle_{c} + 3 \langle \delta^{2} \rangle_{c} 
\langle \delta \rangle_{c} + \langle \delta \rangle^{3}_{c}\\
\langle \delta^{4} \rangle &=&  \langle \delta^{4} \rangle_{c} + 4 \langle \delta^{3} \rangle_{c} 
\langle \delta \rangle_{c} + 3 \langle \delta^{2} \rangle^{2}_{c} \\ \nonumber 
& \hphantom{0}& + 6 \langle \delta^{2} \rangle_{c} \langle \delta \rangle^{2}_{c} + \langle \delta \rangle^{4}_{c} \\
\langle \delta^{5} \rangle &=& \langle \delta^{5} \rangle_{c} + 5 \langle \delta^{4} \rangle_{c} 
\langle \delta \rangle_{c}
+ 10 \langle \delta^{3} \rangle_{c} \langle \delta^{2} \rangle_{c} 
\\ \nonumber && + 10 \langle \delta^{3} \rangle_{c} \langle \delta \rangle^{2}_{c} 
+ 15 \langle \delta^{2} \rangle^{2}_{c} \langle \delta \rangle_{c} \\ \nonumber 
&& + 10 \langle \delta^{2} \rangle_{c} \langle \delta \rangle^{3}_{c} 
+ \langle \delta \rangle^{5}_{c}, 
\end{eqnarray} 
where the subscript $c$ outside the angular brackets denotes a connected moment. Remembering that $\langle \delta \rangle = 0$, these equations simplify to: 
\begin{eqnarray} 
\langle \delta^{2} \rangle &=& \langle \delta^{2} \rangle_{c}  \\
\langle \delta^{3} \rangle &=& \langle \delta^{3} \rangle_{c} \\
\langle \delta^{4} \rangle &=&  \langle \delta^{4} \rangle_{c}  + 3 \langle \delta^{2} \rangle^{2}_{c} \\
\langle \delta^{5} \rangle &=& \langle \delta^{5} \rangle_{c} + 
10 \langle \delta^{3} \rangle_{c} \langle \delta^{2} \rangle_{c}. 
\end{eqnarray} 
Hence, for the second and third order moments, there is no difference in practice between 
the connected and unconnected moments. 

The $N$-point auto-correlation functions, $\xi_N$,  are written in terms of the connected moments: 
\begin{equation}
\xi_N(x_1,\ldots,x_N) = \langle \delta(x_1), \ldots, \delta(x_N)\rangle_c.
\label{eq:autocorr}
\end{equation}

By analogy with the $N$-point auto correlation functions of fluctuations in a single density 
field, we can define the $i+j$-point cross correlation function of two, co-spatial density fields,
with respective density contrasts given by $\delta_1$ and $\delta_2$:
\begin{eqnarray}
\xi_{i,j}(x_1, \ldots, x_i&;&y_1, \ldots y_j) = \\ \nonumber
&&\langle \delta_1(x_1), \ldots, \delta_1(x_i) \,\delta_2(y_1), \ldots, \delta_2(y_j) \rangle_c.
\label{eq:xi}
\end{eqnarray}
In the application in this paper, the first index will refer to the distribution of dark 
matter haloes and the second index to the dark matter. 
When the density contrasts are evaluated at the same spatial location, 
i.e. $x_1=\ldots=x_i=y_1=\ldots=y_j=0$, the connected moments $\xi_{i,j}$ are 
called cumulants of the joint probability distribution function of $\delta_1$ and $\delta_2$ 
(and are sometimes denoted as $k_{i,j}$). 

To generate expressions for the higher order correlation functions of the cross-correlated 
density fluctuations, $\xi_{i,j}$, we will use the method of generating functions (see \S 3.3.3 
of Bernardeau et~al. 2002). A moment generating function is defined 
for the central moments ($\mu_{i,j}$) as a power series in $\delta_1$ and $\delta_2$, which can be 
written as $\chi \equiv \langle \exp \left( \delta_1 t_1 + \delta_2  t_2  \right)\rangle $, 
where $t_1$ and $t_2$ are random variables. This moment generating function can be related to 
the cumulant generating function ($\psi$) for the connected 
cumulants by (see Bernardeau et~al. 2002 
for a proof):  
\begin{equation}
\psi(t_1,t_2) \equiv \ln \chi(t_1,t_2) .
\end{equation}
Then, by taking partial derivatives of $\psi$ and $\chi$ 
evaluated at $t_1=t_2=0$, one can ``generate''  the cumulants and moments:
\begin{eqnarray}
\xi_{i,j}(0) = k_{i,j} &=&  \frac{ \partial^{i+j} } {\partial t_1^i \, \partial t_2^j} \psi \arrowvert_{t_1=t_2=0} \\
\mu_{i,j} &=&  \frac{ \partial^{i+j} } {\partial t_1^i \, \partial t_2^j} \chi \arrowvert_{t_1=t_2=0} = \langle \delta_1^i \delta_2^j \rangle. 
\end{eqnarray}

Following this method we can obtain expressions for the cross-correlation cumulants 
up to order $i+j=5$, grouping terms of the same order:

\begin{eqnarray}
k_{{1,1}} &=& \mu_{{1,1}} 
\label{eq:cumulants1}\\
k_{{2,0}} &=& \mu_{{2,0}} \\ \nonumber 
\\
k_{{3,0}} &=& \mu_{{3,0}} \\
k_{{2,1}} &=& \mu_{{2,1}} \\ \nonumber 
 \\ 
k_{{4,0}} &=& \mu_{{4,0}}-3\,{\mu_{{2,0}}}^{2} \\
k_{{3,1}} &=& \mu_{{3,1}}-3\,\mu_{{2,0}}\mu_{{1,1}} \\
k_{{2,2}} &=& \mu_{{2,2}}-\mu_{{2,0}}\mu_{{0,2}}-2\,{\mu_{{1,1}}}^{2} \\ \nonumber 
\\ 
k_{{5,0}} &=& \mu_{{5,0}}-10\,\mu_{{3,0}}\mu_{{2,0}} \\
k_{{4,1}} &=& \mu_{{4,1}}-4\,\mu_{{3,0}}\mu_{{1,1}}-6\,\mu_{{2,0}}\mu_{{2,1}} \\
k_{{3,2}} &=& \mu_{{3,2}}-\mu_{{3,0}}\mu_{{0,2}}-6\,\mu_{{2,1}}\mu_{{1,1}}-3\,\mu_{{2,0}}\mu_{{1,2}}.
\label{eq:cumulants2}
\end{eqnarray}
Note that these results are symmetric with respect to exchanging the indexes and that we have used the 
fact that $\mu_{1,0}=\mu_{0,1}=0$, since, by construction $\langle \delta_1 \rangle = \langle \delta_2 \rangle = 0$. 

\subsubsection{Higher order correlations: smoothed and discrete density fields}

Sadly, density fluctuations at a point are of little practical use as they cannot be measured reliably, 
as typically we have a finite number of tracers of the density field, i.e. galaxies in a survey 
or dark matter particles in an N-body simulation, and so have a limited resolution view of 
the density field. Furthermore, estimating the $N$-point correlations 
for a modern survey or simulation is time consuming and short cuts are often 
taken, such as restricting the number of configurations of points sampled. 
To overcome both of these problems, moments of the smoothed density field 
can be computed instead of the point moments. 

The smoothed density contrast, $\delta_R$, is a convolution of the density contrast at a point 
with the smoothing window, $W_R$, which has volume $V$:  
\begin{equation}
\delta(x)_R =  \frac{1}{V} \int {\rm d}x^{3'} \delta(x) W_R (x-x'). 
\end{equation}
Typically, the smoothing window is a spherical top-hat in which case $W_R = 1$ for all points 
within distance $R$ from the centre of the window and $W_R=0$ otherwise. 
After smoothing, the cumulants correspond to the $i+j$-point volume-averaged cross correlation 
functions:
\begin{eqnarray}
\bar{\xi}_{i,j}(R) &\equiv&  \int 
{\rm d}^3x_1 \ldots {\rm d}^3x_{i} \quad 
{\rm d}^3y_1 \ldots {\rm d}^3y_{j} \\ \nonumber
&& W_R(x_1)\ldots W_R(x_{i}) \, W_R(y_1)\ldots W_R(y_{j})\xi_{i,j}. 
\label{eq:volxi}
\end{eqnarray}
Eqs. \ref{eq:cumulants1}-\ref{eq:cumulants2} are still valid, with the cumulants replaced by 
volume-averaged cumulants. 

Another issue introduced by the discreteness of the density field is the contribution 
of Poisson noise to the measurements of the cumulants. To take this into account, we 
can modify the moment generating function as follows (\citealt{Peebles1980}):

\begin{eqnarray}
\chi(t_1,t_2)  &=& \langle \exp(f_1 \left( t_1 \right) +f_2 \left( t_2 \right)) \rangle, \\
 f_1  &=& \left( \exp(t_1)- t_1-1 \right) {\bar{n}_1}+ \left( \exp(t_1)-1 \right) \delta_{1}, \\
 f_2  &=& \left( \exp(t_2)- t_2-1 \right) {\bar{n}_2}+ \left( \exp(t_2)-1 \right) \delta_{2}.
\end{eqnarray}
Here, $\bar{n}_{1}$ and $\bar{n}_2$ are the mean number of objects in density field 1 and density field 2 
respectively within spheres of radius $R$. Using this modified generating function, and 
defining  
${\mu}^{\prime}_{i,j} = \langle \left( n_{1} - \bar{n}_{1}\right)^{i} 
\left ( n_{2} - \bar{n}_{2} \right)^{j} \rangle $, 
we obtain the following relations between the volume-averaged, connected $i+j$-point 
cross correlation functions, $\bar{\xi}_{i,j}$, and the central moments, $\mu_{i,j}$:
\begin{eqnarray}
\bar{n}_1^2       \, \bar{\xi}_{2,0} &=&  \mu^{\prime}_{2,0} - \bar{n}_1 \\
\label{eq:cross1}
\bar{n}_1   \bar{n}_2   \, \bar{\xi}_{1,1} &=&  \mu^{\prime}_{1,1} \\
      \bar{n}_2^2 \, \bar{\xi}_{0,2} &=&  \mu^{\prime}_{0,2} - \bar{n}_2 \\ \nonumber 
\\
\bar{n}_1^3       \, \bar{\xi}_{3,0} &=&  \mu^{\prime}_{3,0} + 2 \bar{n}_1 - 3 \mu^{\prime}_{2,0} \\
\bar{n}_1^2 \bar{n}_2   \, \bar{\xi}_{2,1} &=&  \mu^{\prime}_{2,1} - \mu^{\prime}_{1,1}\\
\bar{n}_1   \bar{n}_2^2 \, \bar{\xi}_{1,2} &=&  \mu^{\prime}_{1,2} - \mu^{\prime}_{1,1}\\
      \bar{n}_2^3 \, \bar{\xi}_{0,3} &=&  \mu^{\prime}_{0,3} +  2 \bar{n}_2 - 3 \mu^{\prime}_{0,2} \\ \nonumber 
\\
\bar{n}_1^4       \, \bar{\xi}_{4,0} &=& \mu^{\prime}_{4,0} - 6 \bar{n}_1 + 11 \mu^{\prime}_{2,0} - 6 \mu^{\prime}_{3,0} - 3 \mu^{\prime 2}_{2,0}\\
\bar{n}_1^3 \bar{n}_2   \, \bar{\xi}_{3,1} &=& \mu^{\prime}_{3,1} + 2\mu^{\prime}_{1,1}-3\mu^{\prime}_{2,1}-3\mu^{\prime}_{1,1}\mu^{\prime}_{2,0}\\
\bar{n}_1^2 \bar{n}_2^2 \, \bar{\xi}_{2,2} &=& \mu^{\prime}_{2,2} - \mu^{\prime}_{1,2}- \mu^{\prime}_{2,1} + \mu^{\prime}_{1,1}-\mu^{\prime}_{2,0}\mu^{\prime}_{0,2} \\ \nonumber  
&& - 2 \mu^{\prime 2}_{1,1}\\ 
\bar{n}_1   \bar{n}_2^3 \, \bar{\xi}_{1,3} &=& \mu^{\prime}_{1,3} + 2 \mu^{\prime}_{1,1} - 3 \mu^{\prime}_{1,2} - 3 \mu^{\prime}_{1,1}\mu^{\prime}_{0,2}\\
      \bar{n}_2^4 \, \bar{\xi}_{0,4} &=& \mu^{\prime}_{0,4} - 6 n_2 + 11 \mu^{\prime}_{0,2} - 6 \mu^{\prime}_{0,3} - 3 \mu^{\prime 2}_{0,2}.
\label{eq:cross2}
\end{eqnarray}
Note that these expressions revert to those in the literature for autocorrelation moments in 
the case of either i or j equal to zero (see for instance \citealt{Baugh1995}). Also note that in the 
limit $\bar{n}_1 \rightarrow \infty$, $\bar{n}_2 \rightarrow \infty$,  they correspond to the 
expressions given by Eqs. \ref{eq:cumulants1}-\ref{eq:cumulants2}.

\subsection{Hierarchical amplitudes}

At this point it is useful to define quantities called hierarchical amplitudes which are 
the ratio between the $N$-point, volume-averaged connected moments and the two-point 
volume-averaged connected moment 
raised to the $N-1$ power: 
\begin{equation}
 S_{\rm N} \equiv \frac{ {\bar{\xi}}_{\rm N}} {{\bar{\xi}}_2^{N-1}}. 
\label{eq:sn}
\end{equation}
This form is motivated by the expected properties of a Gaussian field which evolves due 
to gravitational instability (Bernardeau et~al. 2002). 
In the case of small amplitude fluctuations, i.e. on smoothing scales for which 
$\bar{\xi}_{2}(R) \ll 1$, the $S_{\rm N}$ depend only on the local slope of the 
linear perturbation theory power spectrum of density fluctuations and are independent of 
time (\citealt*{JuszkiewiczBouchetColombi1993}; see \citealt{Bernardeau1994} for 
expressions for the $S_{\rm N}$). Similar scalings, but with different values for the $S_{\rm N}$, 
apply 
in the case of distributions of particles which have not arisen through gravitational 
instability, e.g. particles displaced according to Zeldovich approximation 
(see \citealt*{JuszkiewiczBouchetColombi1993}). 

In the case of a Gaussian density field, all of the $S_{\rm N}$ are equal to zero. Initially, 
as perturbations grow through gravitational instability, the two-point connected moment increases. 
The distribution of fluctuations soon starts to deviate from a Gaussian, particularly as 
voids grow in size and cells become empty ($\delta \rightarrow -1$). Voids evolve more slowly 
than overdense regions. There is in principle no limit on how overdense a cell can become. 
As a result, the distribution of overdensities becomes asymmetrical or skewed, with the peak 
of the distribution moving to negative density contrasts and a long tail developing 
to high density contrasts. 
To first order, this deviation from symmetry is quantified by the value 
of $S_{3}$, which is often referred to as the skewness of the density field. Higher order 
moments and hierarchical amplitudes probe progressively further out into the tails of the 
distribution of density contrasts. 

\subsection{Higher order correlations: biased tracers}

We are now in a position to consider the cross-correlation functions for the case of relevance 
in this paper, when the set of objects making up one of the density fields is 
local function of the second density field; the first density field is a biased tracer 
of the second. In our application, one density field is defined by the spatial 
distribution of dark matter haloes and the other by the dark matter. In the case of a local 
bias and small perturbations, the density contrast in the biased tracers ($\delta_1$) 
can be written as an expansion in terms of the underlying dark matter density contrast ($\delta_2$), 
as proposed by  Fry \& Gaztanaga (1993):  

\begin{equation}
\delta_{1}(R) = \sum_{k=0}^\infty \frac{b_k}{k!} \delta_{2}^k(R),
\end{equation}
where the $b_k$ are known as bias coefficients; $b_{1}$ is the linear bias commonly discussed 
in relation to two point correlations. Note that, by construction, we require that 
$\langle \delta \rangle =0$, which implies $b_0 = - \sum_{k=2}^\infty \langle b_k \rangle / k!$. 
The $b_k$, as we shall see later, depend on mass but this is suppressed in our notation.

Using this bias prescription, and following the treatment \cite{FryGaztanaga1993} used for 
autocorrelations, we can write the volume-averaged cross-correlation functions of dark 
matter haloes in terms of the two-point volume averaged correlation function ($\bar{\xi}_{0,2}$) 
and hierarchical amplitudes of the dark matter, $S_{\rm N}$: 

\begin{eqnarray}
{\bar \xi}_{1,1} &=&   b_1   {\bar \xi}_{0,2} + O \left( {\bar \xi}_{0,2}^{2} \right)\\
{\bar \xi}_{2,0} &=&   b_1^2 {\bar \xi}_{0,2} + O \left( {\bar \xi}_{0,2}^{2} \right) \\ \nonumber 
\\
{\bar \xi}_{1,2} &=& b_1   {\bar \xi}_{0,2}^2 \left( c_2+S_3 \right) + O\left( {\bar \xi}_{0,2}^3 \right) \\
{\bar \xi}_{2,1} &=& b_1^2 {\bar \xi}_{0,2}^2 \left( 2\,c_{{2}}+S_{{3}} \right) + O\left( {\bar \xi}_{0,2}^{3} \right)\\
{\bar \xi}_{3,0} &=& b_1^3 {\bar \xi}_{0,2}^2 \left( 3\,c_{{2}}+S_{{3}} \right) + O\left( {\bar \xi}_{0,2}^{3} \right) \\ \nonumber 
\\
{\bar \xi}_{1,3} &=&  b_1   {\bar \xi}_{0,2}^3 \left( 3 S_3 c_2 + S_4 + c_3 \right) +O \left( {\bar \xi}_{0,2}^{4} \right) \\
{\bar \xi}_{2,2} &=&  b_1^2 {\bar \xi}_{0,2}^3 \left( S_4 + 6 S_3c_2+2 c_2^2 + 2 c_3 \right) +O \left( {\bar \xi}_{0,2}^{4} \right)\\
{\bar \xi}_{3,1} &=&  b_1^3 {\bar \xi}_{0,2}^3 \left( 6 c_2^2 + 9 S_3 c_2 + S_4 + 3c_3 \right) +O \left( {\bar \xi}_{0,2}^{4} \right)\\ 
{\bar \xi}_{4,0} &=&  b_1^4 {\bar \xi}_{0,2}^3 \left( 12 c_2^2 + 12 S_3 c_2 + S_4 + 4 c_3 \right)+O \left( {\bar \xi}_{0,2}^{4} \right) \\ \nonumber 
\\
{\bar \xi}_{1,4} &=&  b_1   {\bar \xi}_{0,2}^4  ( 4 c_2 S_4 + 6 c_3 S_3 +  \\ \nonumber 
          &\hphantom{00}&c_4 + S_5 + 3 c_2 S_3^2  ) + O \left( {\bar \xi}_{0,2}^5 \right)\\
{\bar \xi}_{2,3} &=&  b_1^2 {\bar \xi}_{0,2}^4  ( 12 S_3 c_3 + 6 S_3^2 c_2 + 12 S_3 c_2^2 + 6 c_2 c_3 + \\ \nonumber 
          &\hphantom{00}&2 c_4 + S_5 + 8 c_2 S_4 ) + O \left( {\bar \xi}_2^5 \right)\\
{\bar \xi}_{3,2} &=&  b_1^3 {\bar \xi}_{0,2}^4  ( 12 c_2 S_4 + 18 c_3 S_3 + 18 c_2 c_3 + 36 c_2^2 S_3 \\ \nonumber 
          &\hphantom{00}& + 9 c_2 S_3^2 + S_5  + 6 c_2^3 + 3 c_4 ) + O \left( {\bar \xi}_2^5 \right) \\
{\bar \xi}_{4,1} &=&  b_1^4 {\bar \xi}_{0,2}^4 ( 4 c_4 + 24 c_2^3 + S_5 + 72 c_2^2 S_3 + 16 c_2 S_4 + \\ \nonumber 
          &\hphantom{00}& 36 c_2 c_3 + 24 c_3 S_3 + 12 c_2 S_3^2 ) + O \left( {\bar \xi}_2^5 \right)\\
{\bar \xi}_{5,0} &=&  b_1^5 {\bar \xi}_{0,2}^4 ( 20 c_2 S_4 + 15 c_2 S_3^2 + 60 c_2^3 + 30 c_3 S_3 +\\ \nonumber 
          &\hphantom{00}&5 c_4 +120 c_2^2S_3  +  S_5 + 60 c_2 c_3 ) + O \left( {\bar \xi}_2^5 \right)
\end{eqnarray}

\noindent where $c_k = b_k/b_1$. Note it has been shown that these transformations preserve the hierarchical nature of the clustering (\citealt{FryGaztanaga1993}).

\subsection{Numerical Simulations}
To make accurate measurements of the higher order clustering 
of dark matter and dark matter haloes, we use the N-body 
simulations carried out by \cite{Angulo2008}. Two simulation 
specifications were used: i) The BASICC, a high-resolution run 
which used $1448^3$ particles of mass $5.49\times10^{11}\,h^{-1}\,M_\odot$ 
to follow the growth of structure in the dark matter in a periodic 
box of side $1340h^{-1}$Mpc. ii) The L-BASICC ensemble, a suite of 
50 lower resolution runs, which used $448^{3}$ particles of mass 
$1.85\times10^{12}\,h^{-1}\,M_\odot$ in the same box size as the 
BASICC. Each L-BASICC run was evolved from a different realization of the 
initial Gaussian density field. The simulation volume was chosen to allow the 
growth of fluctuations to be modelled accurately on a wide range of scales, including 
that of the baryonic acoustic oscillations (the BASICC acronym stands for 
Baryonic Acoustic oscillation Simulations at the Institute for 
Computational Cosmology). The extremely 
large volume of each box also makes it possible to extract accurate 
measurements of the clustering of massive haloes. The superior mass resolution 
of the BASICC run means that it can resolve the haloes which are predicted to 
host the galaxies expected to be seen in forthcoming galaxy surveys. The L-BASICC 
runs resolve haloes equivalent to group-sized systems. The independence of the 
L-BASICC ensemble runs makes them ideally suited to the assessment of the impact 
of cosmic variance on our clustering measurements. 

In both cases, the same values of the basic cosmological parameters were adopted, 
which are broadly consistent with recent data from the cosmic microwave background and the 
power spectrum of galaxy clustering \citep{Sanchez2006}: the matter density 
parameter, $\Omega_{\rm M}=0.25$, the vacuum energy density parameter, 
$\Omega_{\Lambda}=0.75$, the normalization of density fluctuations, expressed in terms 
of the linear theory amplitude of density fluctuations in spheres of radius $8h^{-1}$Mpc 
at the present day, $\sigma_{8} = 0.9$, the primordial spectral index $n_{\rm s} = 1$, 
the dark energy equation of state, $w=-1$, and the Hubble constant, 
$h=H_{0}/(100{\rm km s}^{-1}{\rm Mpc}^{-1})=0.73$. The simulations were started from 
realizations of a Gaussian density field set up using the Ze'ldovich approximation 
(\citealt{Zeldovich1970}). Particles were perturbed from a glass-like distribution 
(\citealt{Baugh1995,White1994}). The starting redshift for both sets of simulations
was $z=63$. The linear perturbation theory power spectrum used to set up 
the initial density field was generated using the Boltzman 
code {\tt CAMB} (\citealt{Lewis2000}). The initial density field was evolved to 
the present day using a memory efficient version of {\tt GADGET-2} (\citealt{Springel2005}). 

\begin{figure}
\includegraphics[width=8.5cm]{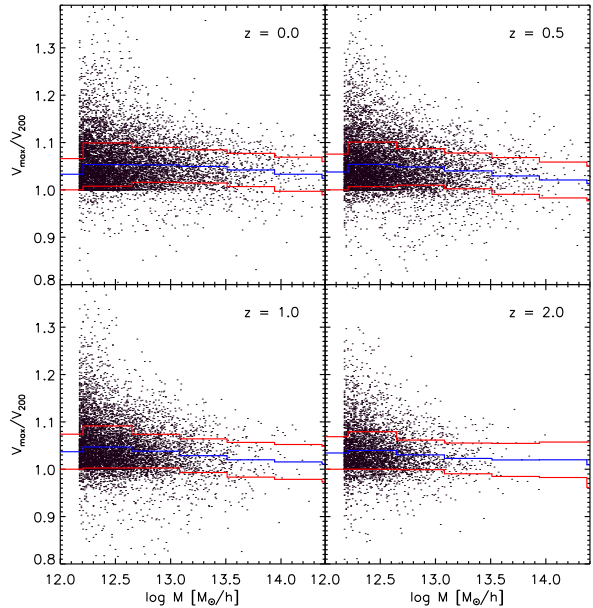}
\caption{ 
The ratio $V_{\rm max}/V_{200}$ as a function of halo mass for gravitationally bound 
haloes in the {\tt BASICC} simulation, which have a minimum of 26 particles. 
$V_{\rm max}$ is the maximum effective circular velocity of the largest substructure within the 
halo and $V_{200}$ is the effective rotation speed at the radius within which the mean 
density is 200 times the critical density, computed using all of the particles within 
this radius.  
Each panel shows the relation at a different redshift as indicated by the legend. 
The red lines show the 20-80 percentile range of the distribution of 
$V_{\rm max}/V_{200}$ values, and the blue lines show the mean. 
}
\label{fig:conc}
\end{figure}

Outputs of the particle positions and velocities were stored from the simulations 
at selected redshifts. Dark matter haloes were identified using a Friends-of-Friends (FOF) 
percolation algorithm (\citealt{Davis1985}) and substructures within these were found 
using a modified version of {\tt SUBFIND} (\citealt{Springel2001}). 
Our default choice is to use the number of particles in a structure as returned by 
the FOF group finder to set the mass of the halo; at the end of Section 3.4 we discuss 
a variation on this to assess the sensitivity of our results to the group finder.
The position of the halo is the position of the most bound particle in the largest 
substructure, as determined by {\tt SUBFIND}.  
In this paper, only gravitationally bound groups with more than 26 particles are considered. 
The {\tt SUBFIND} algorithm also computes several halo properties such as the circular 
velocity profile $V_{\rm c}(r) = (GM(r)/r)^{1/2}$, $V_{\rm max}$, the maximum value of 
$V_{\rm c}$ for the largest substructure, and $V_{200} = V_c(r_{200})$, 
where $r_{200}$ is the radius of a sphere enclosing a volume of mean density 200 
times the critical density. These properties are calculated using only the particles 
which are bound to the main subhalo of the FOF halo; i.e. ignoring all of the other 
substructure haloes within the FOF halo. In the best resolved haloes, substructures 
other than the largest substructure account for at most 15\% of 
the total halo mass \citep{Ghigna1998}. 
Later on in the paper we will present results for the clustering of haloes as a function 
of mass and a second parameter. We have a limited number of output times available to us, so it 
is not feasible to use the formation time of the halo as the second parameter. Instead, we will 
use the ratio $V_{\rm max}/V_{200}$. Fig.~1 shows  $V_{\rm max}/V_{200}$ as a function of halo 
mass at different epochs in the {\tt BASICC} simulation. There is a trend of declining 
$V_{\rm max}/V_{200}$ with increasing halo mass. In cases where the density profile of the 
dark matter halo matches the universal profile advocated by \cite{NFW1997}, $V_{\rm max}/V_{200}$ depends 
on the concentration parameter which characterizes the profile. Haloes in the extreme parts of 
the distribution of $V_{\rm max}/V_{200}$ also have extreme values of the concentration 
parameter (\citealt*{NFW1997}). 
More massive haloes tend to have lower values of the concentration parameter and 
lower values of the velocity ratio $V_{\rm max}/V_{200}$. 
The ratio $V_{\rm max}/V_{200}$ is easier to extract from the simulation, as it 
does not require a parametric form to be fitted to the density profile. There is a correlation 
between formation time and concentration parameter, and hence the ratio $V_{\rm max}/V_{200}$, 
albeit with scatter (\citealt*{NFW1997}; \citealt{Zhao2003}).

\section{Results}

Our ultimate goal is to measure the higher order bias of dark matter haloes. As described 
in Section 2, we follow a novel approach to do this, employing cross moments between haloes and 
the dark matter. The first step in this process is to compute the densities of haloes and dark 
matter on grids of cubical cells of different sizes\footnote{Tests show that density fluctuations 
in cubical cells can be readily translated into counts in spherical cells by simply setting the 
volume of the spherical cell equal to that of the cube. We use cubical cells for speed. The counts 
are regridded to improve the measurement of the rare event tails of the count distribution.}. 
A natural by-product of this procedure is the higher order clustering of the 
dark matter and haloes in terms of the auto-correlation functions. We first present 
the hierarchical amplitudes estimated for the dark matter (\S 3.1) and haloes (\S 3.2) 
using the autocorrelation function higher order moments. In \S 3.3 we show the 
measurements of the cross moments and in \S 3.4 we present the 
interpretation of these results in terms of the bias parameters. 

\subsection{Hierarchical amplitudes for the dark matter}

\begin{figure*} 
\includegraphics[width=16.5cm]{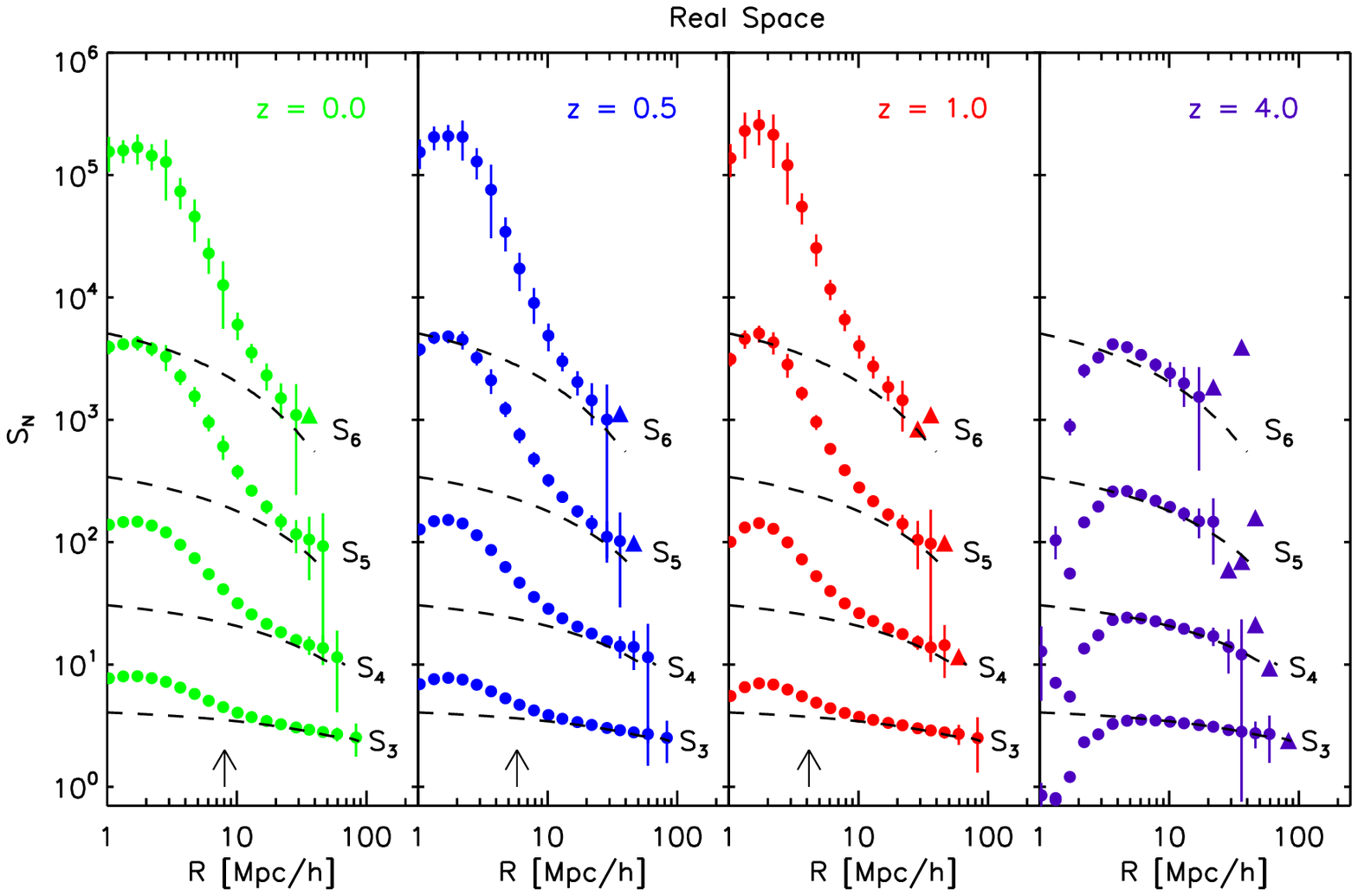}
\includegraphics[width=16.5cm]{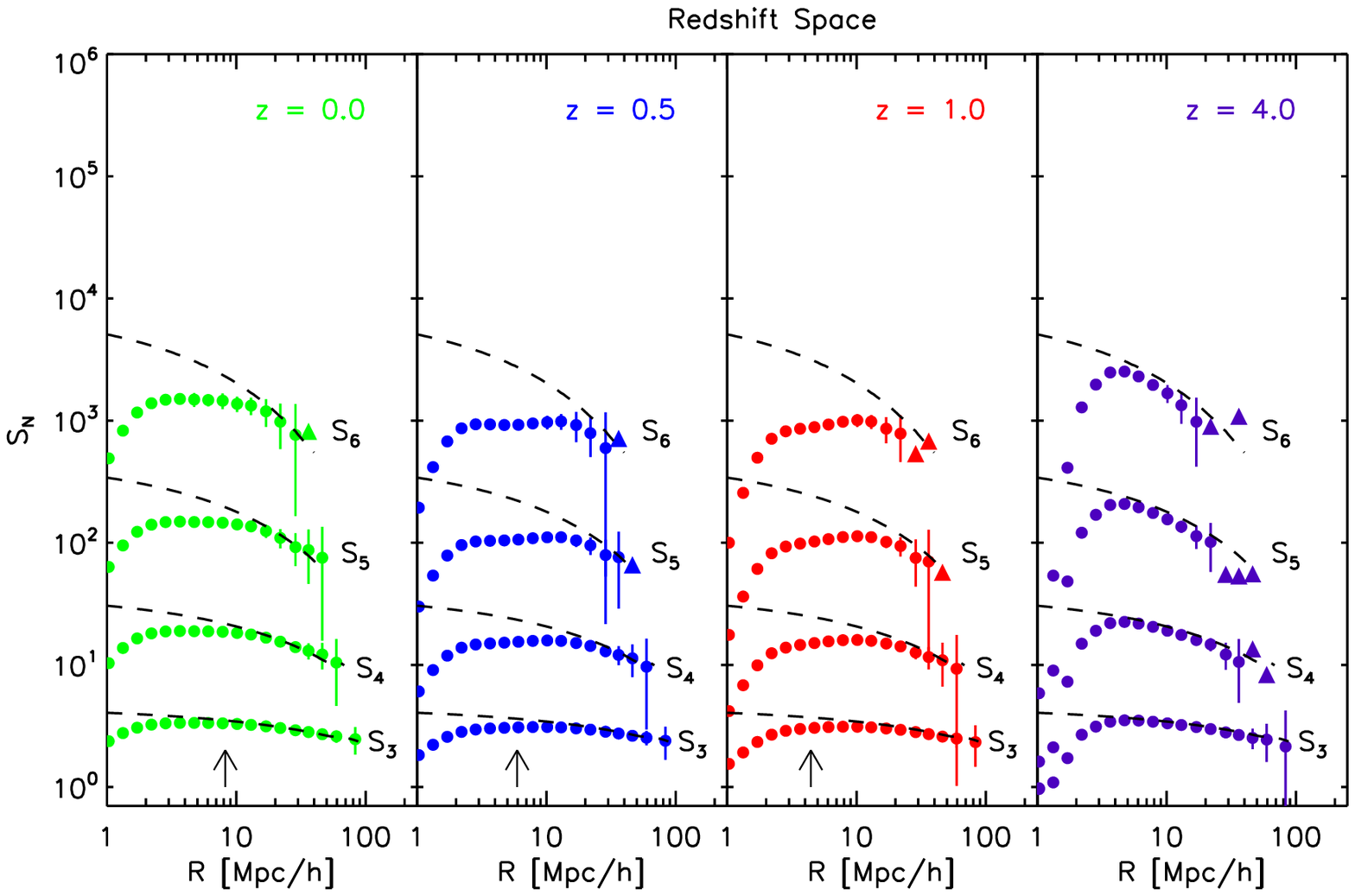}
\caption{ 
The hierarchical amplitudes ($S_{\rm N}$) measured for the dark matter as a function 
smoothing scale, which is plotted in terms of the radius of the sphere with the same 
volume as the cubical cell used.  
The upper panels show the results in real-space and the lower panels show redshift-space. 
Each panel corresponds to a different redshift as indicated by the legend. 
The points show the amplitude for the $S_{\rm N}$ obtained from the {\tt L-BASICC} ensemble, 
after taking the ratio of the median correlation functions, as defined by Eq.~\ref{eq:sn}.
The error bars show the scatter in the measurements over the ensemble, obtained by computing 
$S_{\rm N}$ for each simulation from the ensemble. Error bars are plotted at smoothing scales 
for which the fractional error is less than unity; triangles show scales on 
which the error exceeds unity. 
In both sets of panels, the dashed lines show the 
predictions of perturbation theory in real space (see text for details). Note that no 
correction for shot noise has been applied to the measured amplitudes. 
The arrows indicate the cell radius for which the variance in the counts in cells for the 
dark matter is equal to unity, which is roughly the scale down to which perturbation 
theory should be valid; at $z=4$, this scale is below $R=1 h^{-1}$Mpc. 
}
\label{fig:sp}
\end{figure*}

\begin{figure*} 
\includegraphics[width=15.5cm]{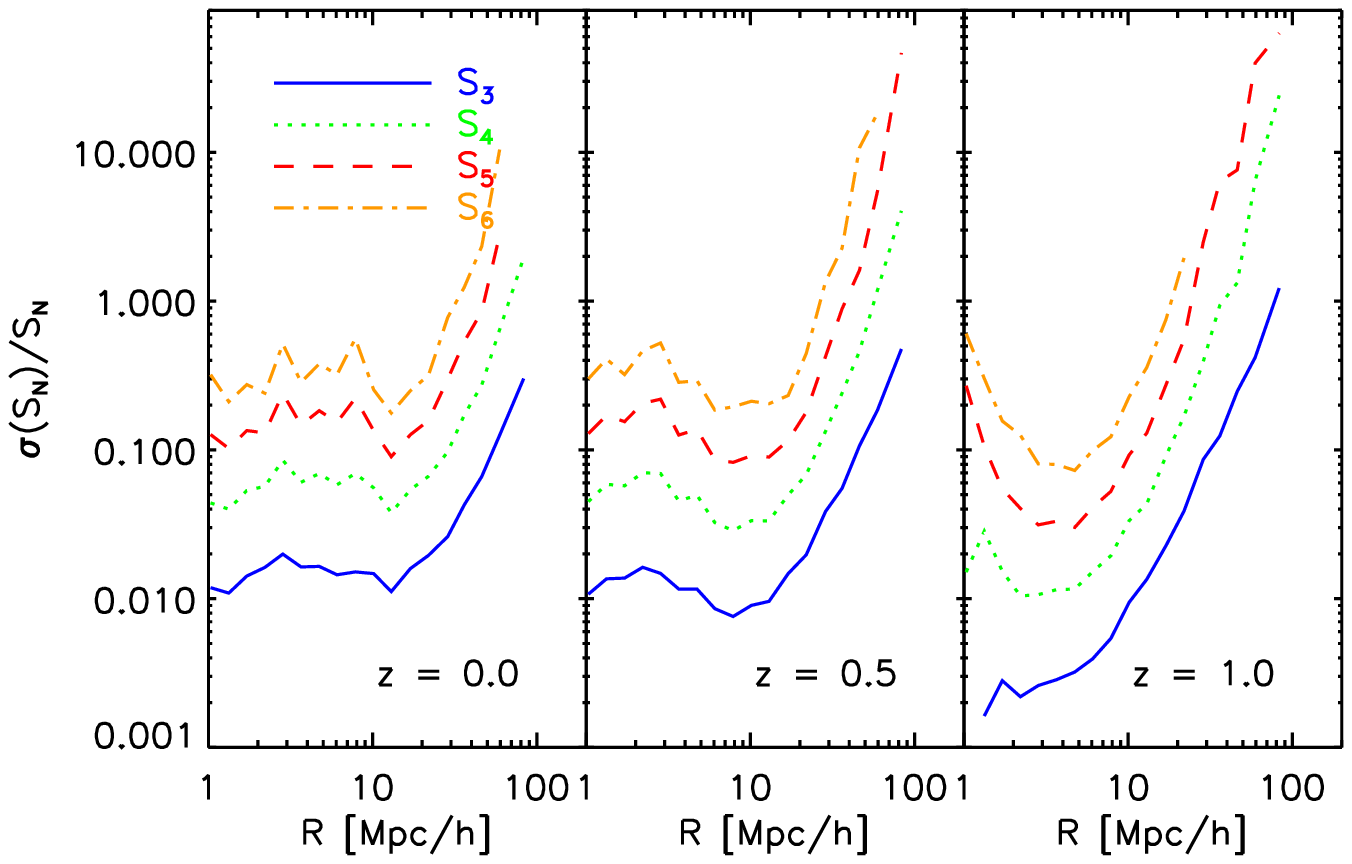}
\caption{
The fractional scatter, $\sigma(S_{\rm N})/S_{\rm N}$, in the 
measured hierarchical amplitudes, as estimated from the 
50 simulations in the L-BASICC ensemble. Different lines 
show the scatter for different orders as indicated by the 
legend. 
}
\label{fig:sperr}
\end{figure*}

Fig.~\ref{fig:sp} shows the hierarchical amplitudes $S_{\rm N}$ measured for the dark 
matter at different redshifts. The upper panels show the results in real-space and the 
lower panels include the effects of redshift space distortions using the distant observer 
approximation. 
The points indicate the median value of the hierarchical amplitudes measured in the 
L-BASICC ensemble and the error bars indicate the variance in these measurements. 
The lines show the hierarchical amplitudes predicted by perturbation 
theory (\citealt{Bernardeau1994}; \citealt*{JuszkiewiczBouchetColombi1993}). 
At the highest redshift plotted, $z=4$, the agreement between the measurements made from the 
simulations and the predictions of perturbation theory is impressive, covering scales 
from $5 h^{-1}$Mpc to $100 h^{-1}$Mpc for $S_3$ and $S_4$. As redshift 
decreases, the simulation results 
for $S_5$ and $S_6$ are slightly higher than the perturbation theory predictions. 
The measurements of $S_3$ from the simulations continue to agree with the perturbation 
theory predictions, but over a narrower range of scales. 
For smoothing scales on which the variance is less than unity, the hierarchical 
amplitudes are expected to be independent of epoch, depending only on the shape of the 
linear perturbation theory power spectrum of density fluctuations (\citealt*{JuszkiewiczBouchetColombi1993}; \citealt{Bernardeau1994, GaztanagaBaugh1995}). 
Fig.~\ref{fig:sp} confirms that this is the case. As the density field evolves, 
the measured hierarchical amplitudes change remarkably little, particularly when 
one bears in mind that the higher order correlation functions change substantially  
between $z=4$ and $z=0$. For example, for a cell of radius $50h^{-1}$Mpc, the 
two-point volume averaged correlation function increases by a factor of $14$ over 
this redshift interval, and the three-point function by a factor of $197$. 

Nevertheless,  
the simulation results do tend to exceed the perturbation theory predictions on all 
scales at all orders as the density fluctuations grow. 

The hierarchical amplitudes measured on small scales differ significantly from the 
predictions of perturbation theory. At $z=4$, the simulation results are below the 
analytical predictions for cell radii smaller than $R \sim 5h^{-1}$Mpc. This behaviour 
is sensitive to the arrangement of particles which is perturbed to set 
up the initial density field. At later times, the memory of the initial conditions is 
erased on small scales and the measured amplitudes greatly exceed the expectations of 
perturbation theory. On these scales, the dominant contribution to the cross correlation 
moments is from particles within common dark matter haloes. 
Note that in Fig.~\ref{fig:sp} we do not correct the measured 
higher order correlation functions for Poisson noise, since the initial density field
was created by perturbing particles distributed in a glass-like configuration 
which is sub-Poissonian. Hence, the dark matter density field is not a random 
sampling of a continuous density field (see \cite{Angulo2008} for an extended 
discussion of this point). 
The turnover in the hierarchical amplitudes seen at small cell radii (e.g. for 
$R<2 h^{-1}$Mpc is due to the finite resolution of the {\tt L-BASICC} simulations; 
the hierarchical amplitudes continue to increase in amplitude on smaller smoothing 
scales in the {\tt BASICC} run.

The lower panels of Fig.~\ref{fig:sp} show the impact of gravitationally induced 
peculiar motions on the hierarchical amplitudes. We model redshift space distortions 
using the distant observer approximation, in which peculiar motions perturb the 
particle position parallel to one of the co-ordinate axes. Virialized structures 
appear elongated when viewed in redshift space. On large scales, coherent bulk flows 
tend to increase the amplitude of correlation functions. There is a modest reduction 
in the amplitude of the hierarchical amplitudes on large scales. On small scales, 
there is a dramatic reduction in the magnitude of the $S_{\rm N}$. The overall 
impact of the redshift space distortions is to greatly reduce the dependence of the 
hierarchical amplitudes on smoothing scale (see \citealt{HoyleSzapudiBaugh2000}). 

The estimated error on the measured hierarchical moments is shown in Fig.~\ref{fig:sperr}, 
in which we plot the fractional error on $S_{\rm N}$ obtained from the scatter in the 
measurements from the L-BASICC ensemble. The plot suggests that the skewness of the dark 
matter can be well measured on all smoothing scales considered from a volume of the size 
of the {\tt L-BASICC} simulation cube. The range of scales over which robust measurements can 
be made of the hierarchical amplitudes becomes progressively narrower with increasing order. 
For example, at $z=0$, reliable measurements of $S_6$ are limited to smoothing radii smaller 
than $R \sim 30h^{-1}$Mpc.

\subsection{The hierarchical amplitudes of dark matter haloes}

\begin{figure}
\includegraphics[width=8.5cm]{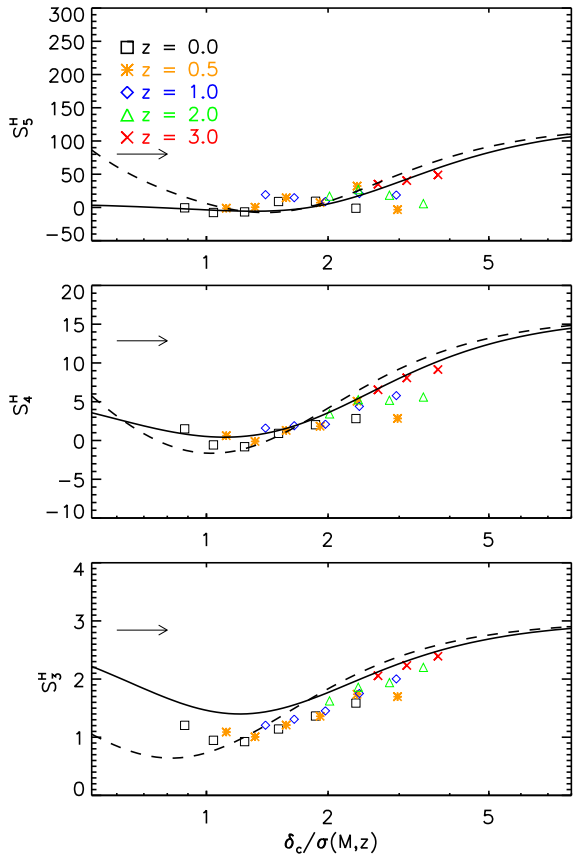}
\caption{ The hierarchical amplitudes of dark matter haloes, plotted as a function of 
the peak height corresponding to the halo mass. In this plot, the hierarchical amplitudes 
are averaged over cell sizes of $20 h^{-1}{\rm Mpc} < R < 50 h^{-1}{\rm Mpc}$. The 
dashed curve shows a theoretical prediction based on the spherical collapse model 
(Mo, Jing \& White 1997) and the solid line shows a revised prediction based upon an 
ellipsoidal collapse, by Sheth, Mo \& Tormen (2001). 
The corresponding hierarchical amplitudes for the dark matter, averaged over the same 
range of cell radii, are indicated in each panel by the arrow. 
}
\label{fig:H}
\end{figure}

The hierarchical amplitudes of dark matter haloes are more complicated than 
those of the dark matter. In addition to a term arising from the evolution of 
the density field under gravitational instability, there is a contribution which 
depends upon height of the peak in the initial density field which collapses to form 
the halo \citep*{MoJingWhite1997}. 
For example, if we consider the second and third order auto-correlation functions 
of haloes given by Eqs.~44 and 47, then the skewness for dark matter haloes, 
$S^{\rm H}_{3}$, is given by: 

\begin{eqnarray} 
S^{\rm H}_{3} & = & \frac{{\bar \xi}_{3,0}}{\left({\bar \xi}_{2,0}\right)^{2}}\\
              & = & \frac{3 b_{2}}{b_{1}^{2}} + \frac{S_{3}}{b_{1}}.
\end{eqnarray}
The gravitational contribution to the skewness, $S_{3}$, 
is diluted by the linear bias factor, $b_{1}$. 
In the case of rare peaks, or, equivalently, haloes with masses far in excess of 
the characteristic mass, $M_{*}$, at a given redshift, $S^{\rm H}_{3}$ approaches 
an asymptotic value. In this limit, $b_{\rm k} \approx b^{k}_{1}$ and so $S^{\rm H}_{3} 
\approx  3$; similar arguments for the fourth and fifth order hierarchical 
amplitudes yield asymptotic 
values of $S^{\rm H}_{4} = 16$ and $S^{\rm H}_{5} = 125$ \citep*{MoJingWhite1997}. 
Massive haloes at high redshift can therefore have non-zero hierarchical amplitudes 
even if the dark matter distribution still has a Gaussian distribution and hence 
$S^{\rm DM}_{\rm p}=0$. 

We plot the hierarchical amplitudes of dark matter haloes in Fig.~\ref{fig:H}, as a function 
of the scaled peak height, $\delta_{c}/\sigma(M,z)$. The simulation results are averaged over smoothing 
radii of $20h^{-1}{\rm Mpc} < R < 50 h^{-1}{\rm Mpc}$. The dashed line shows the prediction  
obtained assuming the mass function of \cite{PressSchechter1974} and the spherical collapse 
model (see \citealt*{MoJingWhite1997}). The solid line shows an improved calculation which uses the 
ellipsoidal collapse model and the mass function derived by \cite{ShethMoTormen2001}. 
There is some dispersion between the simulation results at different redshifts. The 
measurements are in reasonable agreement with the theoretical predictions for large 
values of $\delta_{c}/\sigma(M,z)$. For more modest peaks, the hierarchical amplitudes 
of haloes averaged on large smoothing scales show a dip and are significantly smaller 
than the amplitude recovered for the dark matter. 
The strength of this dip is more pronounced in the measurements from the simulations than 
it is in the theoretical predictions. This discrepancy suggests that the theoretical models 
do not reproduce the trend of bias with halo mass for such objects, as we shall see in \S 3.4.

\subsection{Cross-correlation estimates of higher order clustering }

\begin{figure} 
\includegraphics[width=8.5cm]{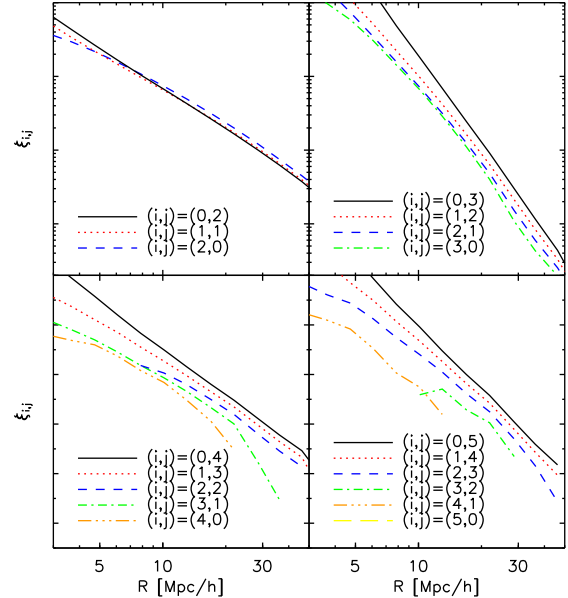}
\caption{ 
Volume-averaged $i+j$-point cross-correlation functions, ${\bar \xi}_{i,j}$, 
measured for haloes of mass $1.1 \times 10^{13} < (M/h^{-1}M_{\odot}) < 2.8 \times 10^{13}$ 
($619 386$ objects) and the dark matter at $z=0$ in the {\tt BASICC} simulation. 
The auto-correlation function of haloes is denoted ${\bar \xi}_{i+j,0}$ and the 
auto-correlation of dark matter by ${\bar \xi}_{0,i+j}$. Each panel shows 
a different order of cross-correlation. The key shows the different 
permutations of cross-correlation function in each case. 
The moments have been corrected for Poisson noise 
due to the finite number of haloes. 
}
\label{fig:cross}
\end{figure}

\begin{figure} 
\includegraphics[width=8.5cm]{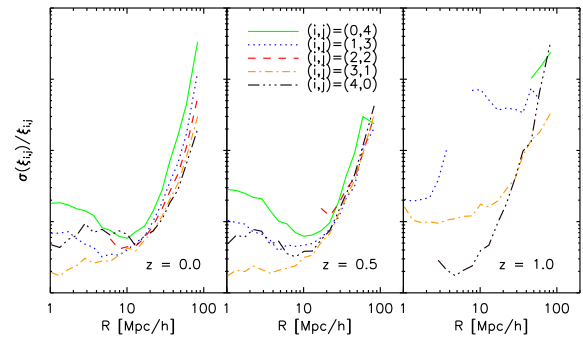}
\caption{
The fractional error on the four-point cross correlation functions, 
estimated from the scatter over the {\tt L-BASICC} runs. 
Each panel shows the results for a different redshift, as shown by the 
key. The legend shows the different permutations of cross-correlation 
moment. To improve the statistics, all the haloes in the {\tt L-BASICC} 
runs have been used in this case.}
\label{fig:crosserror}
\end{figure}

We now switch to estimating cross-correlation functions instead of 
auto-correlation functions. To recap \S 2, to reduce the impact of 
discreteness noise on our measurement of halo clustering, we cross-correlate 
fluctuations in the spatial distribution of haloes with the fluctuation 
in the dark matter density within the same cell. As the order of the correlation 
function increases, the number of possible permutations of halo fluctuations and 
dark matter fluctuations increases. For a given order of correlation 
function, the relation between these permutations can be understood using the 
expressions for the cross-moments given in \S 2.3. The relationship at second 
order is particularly straightforward. The halo autocorrelation function, 
${\bar \xi}_{2,0}$ (recall the first index gives the order of the halo density contrast 
and the second index gives the order of the dark matter density contrast) 
is related to the auto-correlation of the dark matter, ${\bar \xi}_{0,2}$, by 
${\bar \xi}_{2,0} = b_{1}^{2} {\bar \xi}_{0,2}$. The second-order cross correlation 
function, ${\bar \xi}_{1,1}$, is related to the autocorrelation function of dark 
matter by ${\bar \xi}_{1,1}= b_{1} {\bar \xi}_{0,2}$. The primary difference between 
${\bar \xi}_{2,0}$ and ${\bar \xi}_{1,1}$ is therefore a factor of $b_{1}$. This basic 
trend is approximately replicated for any order of correlation function: 
as fluctuations in the halo density are substituted 
by fluctuations in the dark matter, the amplitude of the cross-correlation is 
reduced by a factor which depends on $b_{1}$. Above second order, this factor 
is modulated by higher order bias terms and the hierarchical amplitudes of the 
dark matter (see \S 2.2). The precise relation between the different permutations 
of cross-correlation functions depends upon the values of the bias parameters 
and therefore on the halo mass under consideration. 

We show illustrative examples of volume-averaged cross-correlation functions, 
${\bar \xi}_{i,j}$, 
estimated from the {\tt BASICC} simulation in Fig.~\ref{fig:cross}. Each 
panel shows a different order of clustering, starting with the second moment 
in the top left panel and ending with the fifth order correlation function 
in the bottom right panel. In this plot, the haloes used have masses in 
the range $ 1.1 \times 10^{13} < (M_{\rm halo}/h^{-1}M_{\odot}) <  2.8 \times 10^{13}$ 
and the clustering is measured at $z=0$. The top-left panel of Fig.~\ref{fig:cross} shows that 
there is little difference in the amplitude of the second-order correlation 
function on large smoothing scales between the different permutations 
of $i,j$. This implies that for these haloes, the linear bias term 
$b_{1} \approx 1$. The correlation functions are, however, different on 
small scales. The autocorrelation function of the dark matter (${\bar \xi}_{0,2}$) is steeper 
than the autocorrelation of haloes (${\bar \xi}_{0,2}$). 
The cross-correlation functions are different on large scales for third, fourth 
and fifth orders. The difference in amplitude is fairly independent of scale for 
cells with radii $R>10 h^{-1}$Mpc. Since the linear bias of this sample of haloes 
is close to unity, this difference is driven by the higher order bias terms and 
the hierarchical amplitudes of the dark matter. 
We plan to model the full behaviour of the cross-correlation functions, including the 
small scale form, using the halo model in a future paper. 

One might be concerned that replacing fluctuations in halo density by fluctuations 
in dark matter in the higher order correlation functions leads to a reduction in the 
clustering amplitude (as is indeed apparent in Fig.~\ref{fig:cross}). However, 
this is more than offset by a reduction in the noise or scatter of the measurement. 
The fractional error on the measurements of the cross-correlation functions is 
plotted in Fig.~\ref{fig:crosserror}.
The scatter is estimated using the {\tt L-BASICC} 
ensemble. Each panel shows the scatter at a different redshift. 
The cross correlation ${\bar \xi}_{1, i+j-1}$ (i.e. one part halo fluctuation, $i+j-1$ parts 
dark matter fluctuation) gives the optimal error estimate, with a performance 
comparable to the auto-correlation of the dark matter.
At $z=1$, it is not possible to measure the four-point autocorrelation function 
of this sample of haloes, even with a box of the size of the {\tt L-BASICC} runs. 
Nevertheless, it is possible to measure the bias factors relating the four-point 
functions of haloes and mass using the cross-correlation. 
Our use of a cross-correlation estimator therefore allows us to extend the measurements 
of the higher order clustering of haloes to orders and redshifts that would not be 
possible using auto-correlations.

\begin{figure*}
\includegraphics[width=8.7cm]{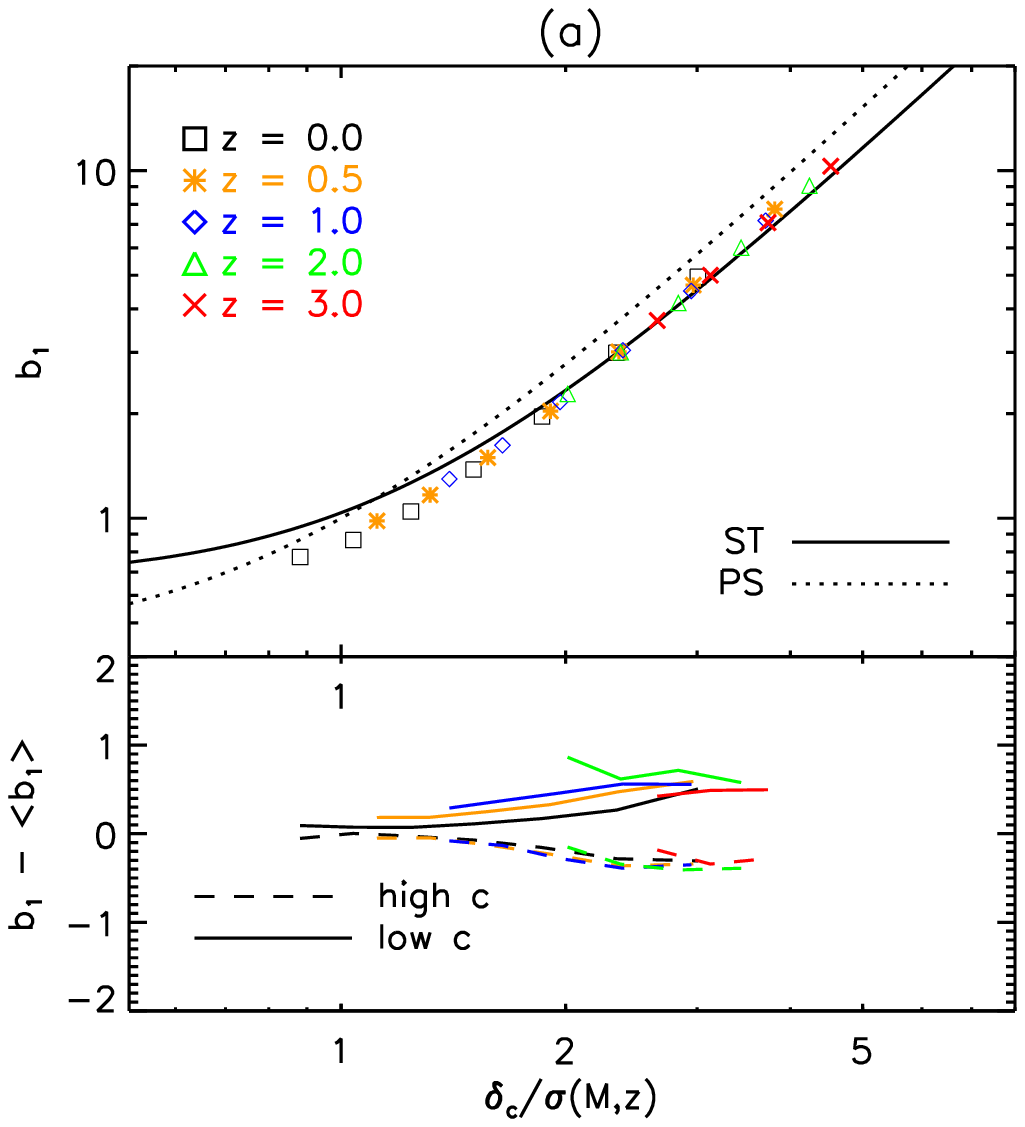}
\includegraphics[width=8.7cm]{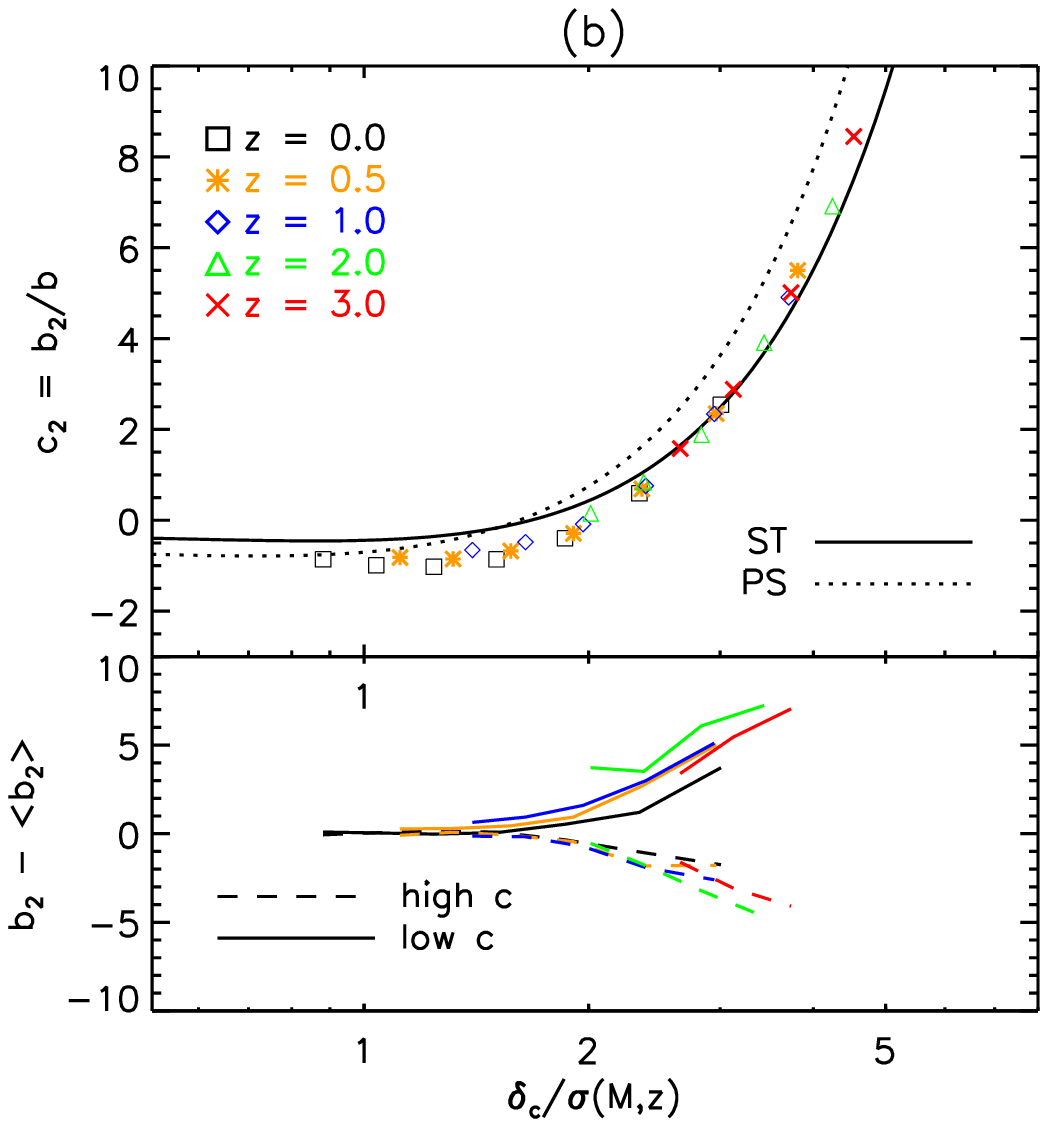}
\includegraphics[width=8.7cm]{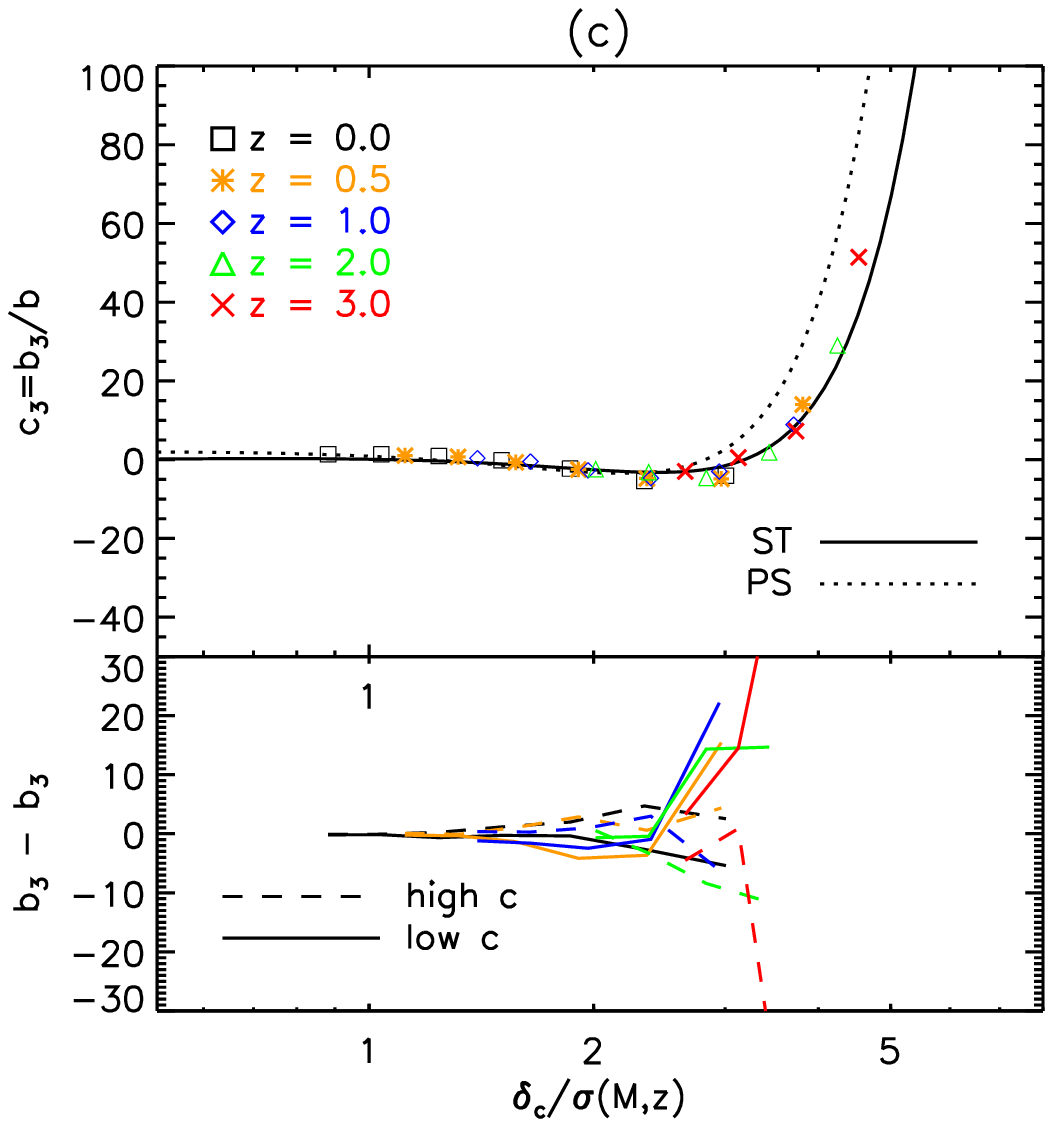}
\includegraphics[width=8.7cm]{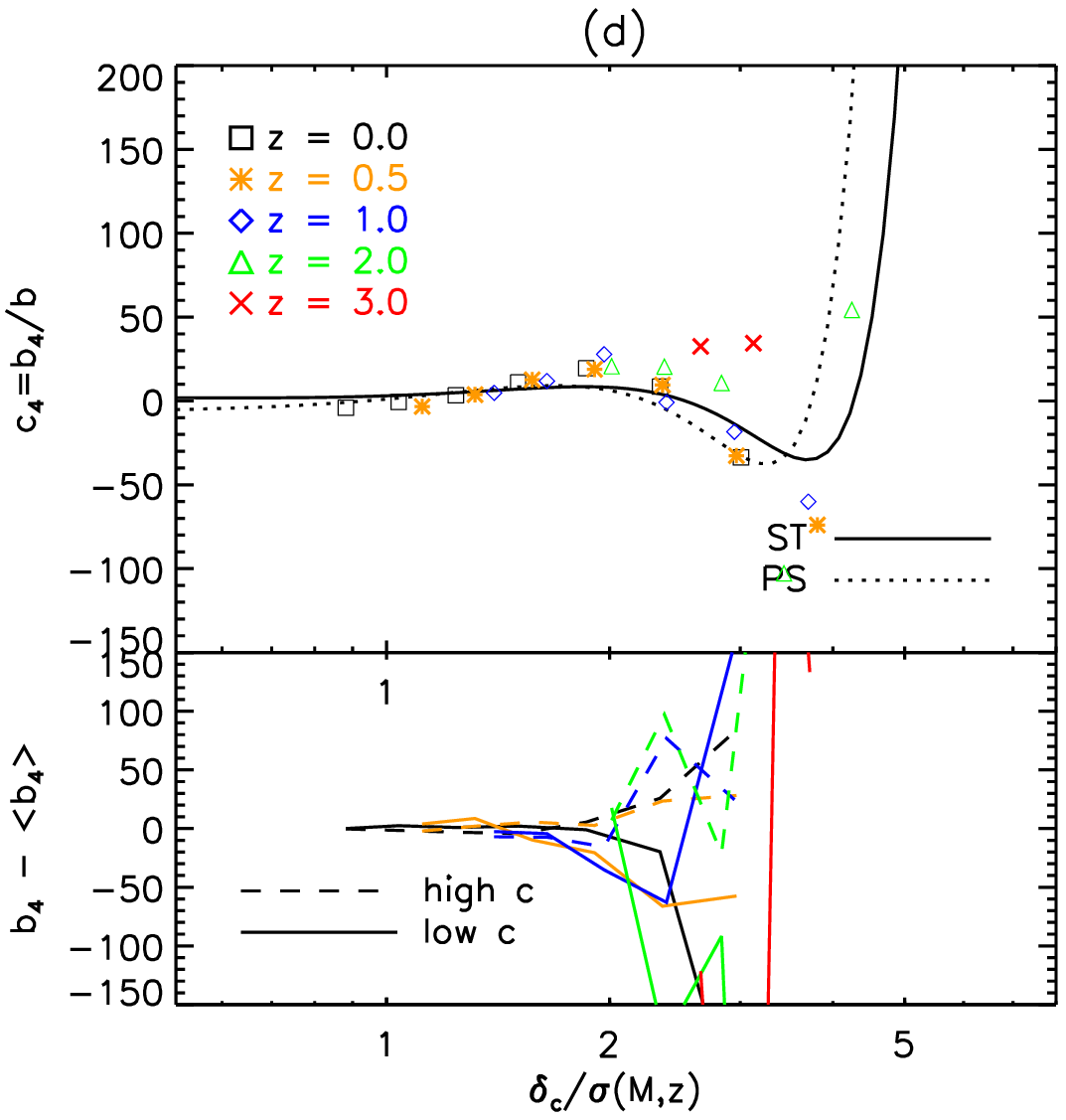}
\caption{
The bias parameters as a function of halo mass parametrized by $\nu = \delta_{\rm c}/\sigma(M,z)$.
Each plot shows a different order of bias parameter: a) linear bias $b_1$, 
b) the ratio of the 2nd order bias, $b_2 / b_1$, c) the ratio of the 3rd order bias, $b_3 / b_1$ 
and d) $b_4/b_1$. 
In the lower panel of each plot, the residual bias parameters for the 20\% of haloes with the highest or lowest 
values of $V_{\rm max}/V_{200}$, a proxy for concentration, are plotted. In the upper panels, symbols 
show the measurements for different output redshifts, as indicated by the key. The same line colours 
are used to show the results for different redshifts in the lower panels. 
In the upper panel of each plot, we plot two theoretical predictions for the bias parameters, given 
by Mo, Jing \& White (1997) and Scoccimarro et~al. (2001). 
}
\label{fig:bias}
\end{figure*}

\subsection{The bias parameters of dark matter haloes}

We now use the cross correlation functions to estimate the linear and higher order 
bias parameters of dark matter haloes. As we demonstrated in the previous section, 
the best possible measurement of the $i+j^{\rm th}$ order correlation function is 
obtained when the cross correlation function is made up of one part fluctuation in 
halo density and $i+j-1$ parts dark matter fluctuation: i.e. in our notation 
${\bar \xi}_{1,i+j-1}$. 
This approach, combined with the huge volume of our simulation, makes it possible, 
for the first time, to measure the third and fourth order bias parameters, and to do 
so using narrow mass bins.  

In this section, we use the higher resolution {\tt BASICC} run, which can resolve 
the largest dynamic range in halo mass. 
We use the higher order correlation function measurements over the range of smoothing 
radii $ 15 < (R/h^{-1}{\rm Mpc}) < 50$ to estimate the halo bias parameters. 
The large volume of the {\tt BASICC} simulation means that we can 
make robust measurements of the higher order correlation functions out to larger smoothing 
radii than is possible with the smaller Millennium simulation. The smallest scale we use 
is set by the requirement that the expansion relating the overdensity in haloes 
to the overdensity in dark matter (Eq. 44) is a good approximation, i.e. when ${\bar \xi} \ll 1$. The scales 
we use to extract the halo bias parameters are considerably larger than those 
Gao, Springel \& White (2005) and Gao \& White (2007) were able to use in 
the Millennium. We use the simulation outputs at redshifts of 
$z=0, 0.5, 1, 2$ and 3 to measure the clustering of haloes.  

The results for the first, second, third and fourth order bias parameters of dark matter 
haloes are presented in Fig.~\ref{fig:bias}. Each panel corresponds to a different order. 
The upper half of each panel shows the respective order of bias parameter as a function 
of halo mass, expressed in terms of the peak height corresponding to the halo mass, 
$\delta_{\rm c}/\sigma(M,z)$. The lower half of each panel shows the deviation from 
the bias parameter extracted for a given mass for samples of the 20\% of haloes in the 
mass bin with the highest and lowest values of $V_{\rm max}/V_{200}$, which we are using 
as a proxy for halo concentration. Different symbols in the upper panels show the measurements 
at different output redshifts in the {\tt BASICC} run, as indicated by the key; the same 
colours are used to draw the lines showing results for samples defined by 
different $V_{\rm max}/V_{200}$ values at the same output redshifts in the lower panels. 

In Fig.~\ref{fig:bias}, there is remarkably little scatter between the results obtained 
from the different output redshifts for the case of the overall bias as a function of mass. 
This is encouraging, as it shows that our results are not affected by resolution (haloes 
with similar values of $\delta_{\rm c}/\sigma(M,z)$ at different output times are made up 
of very different numbers of particles). Gao, Springel \& White (2005) were able to measure 
the linear bias parameter up to haloes corresponding to peak heights of 
$3 \sigma$; we are able to extract measurements for 
haloes corresponding to $5 \sigma$ peaks. 

In the upper sub-panels of Fig.~\ref{fig:bias}, we show two theoretical predictions for 
the bias parameters of dark matter haloes. The dotted lines show the predictions from 
\cite*{MoJingWhite1997}, based on an extension of Press \& Schechter's 
(1974)\nocite{PressSchechter1974} theory for 
abundance of dark matter haloes and the spherical collapse model. The solid lines show 
the calculation from \cite{Scoccimarro2001} which uses the mass function of \cite{ShethTormen1999}.
Our results tend to agree best with the later, although the measurements favour a steeper dependence of bias on 
peak height at all orders. For less rare peaks, neither theoretical model gives a 
particularly good fit to the simulation results. A similar trend, albeit with more 
scatter between the results at different output redshifts was found by \cite*{GaoSpringelWhite2005} 
(see also \citealt{Wechsler2006}; \citealt*{JingSutoMo2007}). 

Previous studies have reported a dependence of clustering strength on a second halo 
property besides mass, such as halo formation time or concentration (\citealt{Wechsler2006}; 
Gao \& White 2007). We do not have sufficient output times to make a robust estimate of formation 
time so we use a proxy for halo concentration instead, $V_{\rm max}/V_{200}$. 
We find that the clustering of high peak haloes is sensitive to whether the halo has 
a high or low value of $V_{\rm max}/V_{200}$. The 20\% of haloes with the lowest values 
of $V_{\rm max}/V_{200}$ within a given mass bin (i.e. those with the lowest 
concentrations) have the largest linear and second order bias terms. This result agrees 
with previous estimates of the dependence of the linear bias term on halo concentration 
(\citealt{Wechsler2006}). 
 
The peak height dependence of the third and fourth order bias terms for haloes split by 
$V_{\rm max}/V_{200}$ is more complicated. Fig.~\ref{fig:bias} shows that the third order 
bias depends on our concentration proxy in a non-monotonic fashion. The trend for the fourth 
order bias is reversed compared with the results for the first and second order bias parameters: 
low concentration haloes have a negative value of the fourth order bias. We note that it would 
not be possible to measure a fourth order bias at all using halo auto-correlation functions. 

One might be concerned that our results could be sensitive to the operation of the group finder. 
In particular, it is well known that the FoF algorithm can sometimes spuriously link together 
distinct haloes into a larger halo, through bridges of particles (e.g. \citealt{ColeLacey1996}). 
We therefore carried out the exercise of relabelling the mass of each halo by the mass of the 
largest substructure as determined by {\tt SUBFIND}. In the rare cases in which haloes are 
incorrectly linked into a larger structure, using instead the {\tt SUBFIND} mass would result 
in a significant shift in the mass bin to which the halo is assigned. Moreover, one would 
expect that low concentration FoF haloes would be more prone to being broken up in this way. 
However, we found no change in our results upon following this procedure, demonstrating that 
the trends we find for the dependence of bias on mass and concentration are robust.

\section{Summary and Discussion}

In this paper, we have combined ultra-large volume cosmological simulations with a novel 
approach to estimating the higher order correlation functions of dilute samples of objects. 
The large simulation volume allows us to extract bias parameters on large scales, which 
follow linear perturbation theory more closely, and provides us with large samples of high 
mass haloes from which robust clustering measurements can be made. The cross-moment 
counts-in-cells technique we use to estimate the higher order clustering of dark matter 
haloes has superior noise performance to traditional autocorrelation functions, allowing 
us to probe clustering to higher orders.  These improvements made it possible to extend 
previous work on the assembly bias of dark matter haloes in a number of ways. We have been 
able to extract measurements of halo clustering for objects corresponding to $5 \sigma$ peaks, 
almost twice as high as in earlier studies. We have also presented, for the first time, 
estimates of the higher order bias parameters of haloes, up to fourth order, and using 
narrow mass bins.  

Our results are in qualitative agreement with those in the literature where they overlap. 
We find that the linear bias factor, $b_1$, is a strong function of mass, varying by an order 
of magnitude for peaks ranging in height from $\delta_{c}/\sigma(M,z)=1$ to $5$. We use the 
ratio of the maximum of the effective halo rotation speed to the speed at the virial radius, 
$V_{\rm max}/V_{200}$ as a proxy for halo concentration. High mass, high $V_{\rm max}/V_{200}$
haloes are less strongly clustered than the same mass haloes with low values of 
$V_{\rm max}/V_{200}$; haloes with $\delta_{c}/\sigma(M,z) \sim 4$ display second order 
clustering that differs by $\approx 25 \%$ between the 20\% with the lowest values 
of  $V_{\rm max}/V_{200}$ and the 20\% of the population with the highest values of this ratio.

It is reassuring that we recover a similar dependence of the linear bias on halo 
mass when labelling haloes by $V_{\rm max}/V_{200}$ as other authors found using 
the concentration parameter (\citealt{Wechsler2006}). This trend 
is the opposite to that recovered when halo samples are split by formation time. 
\cite*{GaoSpringelWhite2005} found no dependence of the clustering signal on halo 
formation time for massive haloes. This is puzzling since formation time and 
concentration are correlated, albeit with scatter (e.g. \citealt{Neto2007}). 
\cite*{CrotonGaoWhite2007} have argued that this suggests that an as yet unknown 
halo property is a more fundamental property in terms of determining the clustering 
strength (for theoretical explanations of the physical basis of assembly bias 
see e.g. \citealt{Zentner2007,Keselman2007,Dalal2008} ). 

The second order bias parameter, $b_2$, displays qualitatively similar dependences on mass and 
$V_{\rm max}/V_{200}$ to $b_1$ with the difference that $b_2$ is negative around 
$\delta_{c}/\sigma(M,z) \sim 1$. The third and fourth order bias parameters are more 
complicated, being essentially independent of mass until peaks 
$\delta_{c}/\sigma(M,z) \sim 2-3$ are reached, where there is a dip in bias before 
a rapid increase for rarer peaks. The dependence on $V_{\rm max}/V_{200}$ is also different 
at third and fourth order. 

We compared our measurements for the bias parameters with analytic predictions. For haloes 
corresponding to rare peaks, the trend in linear bias versus peak height is intermediate 
between the predictions of \cite*{MoJingWhite1997}, which are based on Press \& Schechter's 
(1974)\nocite{PressSchechter1974} theory for the abundance of haloes and the 
spherical collapse model, and the calculation of Sheth, Mo \& Tormen (2001) and 
Scoccimarro et al. (2001), based on ellipsoidal collapse and an improved estimate of 
the halo mass function. Both analytic calculations predict a weaker dependence of $b_1$ on 
peak height around $\delta_{c}/\sigma(M,z) \sim 1$ than we find in the simulation. 
The comparison between the simulation measurements and the analytic predictions is similar 
for $b_2$. For the third and fourth order bias parameters, the simulation results are in good 
agreement with the analytic predictions for modest peaks. For rare peaks, the bias parameters 
measured from the similar are again in between the two analytic predictions. 

Observations of clustering are already entering the regime in which our simulation 
can play an important role in interpreting the measurements. Existing observations 
of high redshift quasar clustering suggesting that these objects live in haloes 
corresponding to $\sim 5-6$ sigma peaks in the matter distribution at $z=4$ (White, 
Martini \& Cohn 2007). 
Future galaxy surveys, due to the volume covered and number of galaxies targeted, will yield 
measurements of clustering with unprecedented accuracy, to higher orders than the two-point 
function. The measurements presented in this paper will provide invaluable input to future 
models of galaxy clustering based on halo occupation distribution models, which have been 
modified such that galaxy clustering is a function of mass and a second halo property. 

\section*{acknowledgements} 

We acknowledge Liang Gao, Shaun Cole and Carlos Frenk for helpful discussions 
and Lydia Heck for managing the Cosmology Machine at Durham which was used to 
run the simulations used in this paper. We also thank Robert Smith, Martin White 
and Andrew Zentner for useful comments on the preprint version of this paper. 
REA is supported by a PPARC/British Petroleum sponsored Dorothy Hodgkin postgraduate 
award. CMB is funded by a Royal Society University Research Fellowship. This work was 
supported in part by a rolling grant from STFC.

\bibliographystyle{mn2e}
\bibliography{bib}

\label{lastpage}
\end{document}